\documentclass[conference,comsoc,10pt]{IEEEtran}
\IEEEoverridecommandlockouts
\usepackage{cite}
\usepackage{amsmath,amssymb,amsfonts}
\usepackage{algorithmic}
\usepackage{graphicx}
\usepackage{textcomp}
\usepackage{xcolor}
\usepackage{booktabs}
\def\BibTeX{{\rm B\kern-.05em{\sc i\kern-.025em b}\kern-.08em
    T\kern-.1667em\lower.7ex\hbox{E}\kern-.125emX}}
\usepackage{bbm}
\usepackage{array}
\newcolumntype{L}[1]{>{\raggedright\let\newline\\\arraybackslash\hspace{0pt}}m{#1}}
\newcolumntype{C}[1]{>{\centering\let\newline\\\arraybackslash\hspace{0pt}}m{#1}}
\newcolumntype{R}[1]{>{\raggedleft\let\newline\\\arraybackslash\hspace{0pt}}m{#1}}
\usepackage{comment}

\begin{document}

\title{Analysis of LoRaWAN Uplink with Multiple Demodulating Paths and Capture Effect
}

%{\title{{Lower Bounds for the Frame Reception Success Probability in the LoRaWAN Uplink}}}
%{\title{\nr{Analysis of Frame Successful Reception  Probability in LoRa-based Communications}}}
%\title{Analysis of packet reception success probability in LoRa-based communications}

%{\title{{Approximation of the Frame Reception Success Probability in the LoRaWAN Uplink}}}
\author{
\IEEEauthorblockN{Ren\'{e} Brandborg S{\o}rensen,  \IEEEmembership{Student Member, ̃IEEE,} Nasrin Razmi, %\IEEEmembership{Student Member, ̃IEEE,} 
\\ Jimmy Jessen Nielsen, \IEEEmembership{Member, ̃IEEE,} Petar Popovski,~\IEEEmembership{Fellow, ̃IEEE}}
\thanks{This work has been supported by the European Research Council (ERC Consolidator Grant Nr. 648382 WILLOW) within the Horizon 2020 Program.}
\thanks{All authors are with the Department of Electronic Systems, Aalborg University, Denmark (Email: \{rbs,razmi,jjn,petarp\}@es.aau.dk).}
}

\maketitle

\begin{abstract}

%The probability of successfully transmitting frame based on LoRa is
%evaluate --> model and analyze
%Low power wide area (LPWA) networks have attracted considerable attention for modeling and standardizing in the academic and industry research aspects. Scalability analysis is one of the most important points in this technology.
%The probability of successful reception (PSR) a frame based on LoRa is of great interest to industry and academia working with the popular LoRa-based LoRaWAN protocol.} In this paper, we \nr{model and analyze} the PSR of a LoRa frame in a \nr{real scenario of} LoRaWAN cell taking into account non-orthogonality of the spreading factors (SFs) in LoRa and the timing of any collisions, which has a great impact on demodulation success in the LoRa hardware. \nr{We also consider and evaluate uniform, distance and Load equivalent policies for SF allocation
%Standardisation of \nr{Long Range} Wide Area Networks (LPWANs) such
Low power wide area networks (LPWANs), such as the ones based on the LoRaWAN protocol, are seen as enablers of large number of IoT applications and services. In this work, we assess the scalability of LoRaWAN by analyzing the frame success probability (FSP) of a LoRa frame while taking into account the capture effect and the number of parallel demodulation paths of the receiving gateway.
We have based our model on the commonly used {SX1301 gateway chipset}, which is capable of demodulating {up to} eight frames simultaneously; however, the results of the model can be generalized to architectures with arbitrary number of demodulation paths. We have also introduced and investigated {three} policies for Spreading Factor (SF) allocation. Each policy is evaluated in terms of coverage {probability}, {FSP}, and  {throughput}. The overall conclusion is that the presence of multiple demodulation paths introduces a significant change in the analysis and performance of the LoRa random access schemes. 
 %\nr{The results show that the number of modulation paths has an important effect on the PSR. }
\end{abstract}
%\nr{ I have changed the "cell" to "network" because the LoRa is not cellular based. }

\begin{IEEEkeywords}
LoRa, LoRaWAN, {frame{ {s}uccess} {p}robability (FSP)}, Collision Timing, Capture-effect, SX1301{.}
\end{IEEEkeywords}

\section{Introduction} \label{sec:intro}
LoRaWAN is a popular {low power wide area network (LPWAN) protocol.} %rbs: I think LPWAN is more accurate and equally descriptive
%Long Range Wide Area Network (LoRaWAN) is a popular protocol \nr{to provide long range communication with low energy consumption.} 
It is based on Semtechs' proprietary LoRa modulation, which uses a chirp spread spectrum {(CSS)} to enable long range and resilient transmissions. The access scheme of LoRa{WAN resembles a pure Aloha protocol, {however LoRaWAN provides several} narrow-band channels and quasi-orthogonal spreading factors {(SF)} {to make LoRaWAN} scalable.}
%but it can provide high scalability by using different carrier frequency (CF), bandwidth (BW), spreading factor (SF) and coding rate (CR). 
A considerable research interest has arisen recently about the obtainable link budget of LoRa transmissions, along with studies of capture effect, collisions, and the non-orthogonality of {SFs} in LoRa-based networks.

{There are three mechanisms in LoRaWAN that determine the network performance: coverage, capture effect, and the demodulation capabilities of the receiver. All of these are affected by the {SF} allocation.}
%spreading factor 
The coverage probability is determined by the device location and the sensitivity of the {SF}. The capture probability is a function of the interference from transmissions using the same SF (co-SF interference) or different SFs (inter-SF interference). Finally the choice of the SF controls the tradeoff between the transmission robustness and collision probability. 
On one hand, the transmissions that use lower SFs are less robust to noise and interference, resulting in a worsened coverage. On the other hand, lower SF results in a higher rate, which, for fixed payload, translates into a shorter packet duration. This implies that the probability to cause or experience collision (interference) for that packet is lowered for lower SFs. In addition, a transmission with a lower SF occupies a demodulation path in the demodulator for a shorter time interval compared to a transmission with a larger SF.
%\nr{In the context of performance evaluation in LoRaWAN, co-SF and inter-SF interferences have a critical effect. Co-SF and inter-SF interferences are caused by the same channel transmissions with using the same and different SFs, respectively.} Coverage, inter-SF interference and demodulation capabilities of the receiver are affected by SFs. {Transmissions using lower SFs are less robust to \nr{the} noise and interference} and thus {have} worse coverage at \nr{long}er distances. \nr{The transmission rate is higher for the lower SFs, therefore} it \nr{causes} \nr{insignificant} interference to the other transmissions {and \nr{ also experience lower interference}.} In the demodulator, a {lower SF} transmission only takes up a demodulation path for the very short period of the transmission. 

%{\pp{I am putting this paragraph here as it is explanatory how LoRA works and what are the tradeoffs. It needs to be improved, by defining what co-SF and inter-SF interference means. Then you can proceed to literature review.}

%There are  three mechanisms in LoRaWAN that determine the network performance, and they are all impacted by Spreading Factor (SF) allocation. The coverage, inter-SF interference and the demodulation capabilities of the receiver. %A SF7 
 %In the case of \nr{ the higher SF} transmission\nr{,} the opposite statements are true.}
%\nr {In this paper, we evaluate the packet successful probability (PSP) of LoRaWAN under the real scenario. The timing has an important effect on the PSP.    }

%

The initial research in LoRaWAN modeled the protocol as pure Aloha channels for each {SF} and channel pair {as in} \cite{8030482}. This work neglects {inter-SF interference and capture effect.} 
%This work neglects the capture effect and assumes that the SFs are perfectly orthogonal.
%The coverage of the uplink of a single LoRaWAN cell has been investigated in \cite{7803607,7996381,8268120,  8430542,8480649}.
{ The performance of LoRa} has been analyzed and modeled for a single LoRaWAN cell using stochastic geometry and considering capture effect in \cite{7803607,7996381,8268120,  8430542,8480649}. 
 %using stochastic geometry and by taking capture effect into account.
The authors in \cite{7803607} model the co-SF interference by considering the interference from the strongest co-SF interferer device. The results show that, the co-SF interference effect increases as the number of devices increases, such that the network becomes interference-limited. The authors in \cite{7996381} provide a framework to evaluate only the co-SF interference by considering the level of overlap between the interfering packets. 
%\rbs{Only} co-SF interference has been \rbs{considered} in \cite{7803607} and \cite{7996381}, where model the strongest co-SF interferer and account for overlaps in both time and frequency, \nr{respectively}. \rbs{obs! previous sentence makes no sense}
Inter-SF interference is modeled with the Co-SF interference in \cite{8268120, 8430542, 8480649}. The authors of \cite{8268120} consider the allocation of SFs in order to maximize the average coverage probability by {assigning SFs to devices based on {their} distance to the {gateway} and SF sensitivities.} 
% \rbs{assigning SFs to devices based on the distance to the AP and SF sensitivity.} 
%\pp{determining SF regions -PP: What does it mean?}
The scalability and throughput of LoRaWAN deployments have been evaluated in \cite{8430542,8480649} based on the capture effect and coverage models. Both papers verify that both co-SF and inter-SF transmissions have a considerable effect on the capture probability, and thereby on the LoRaWAN scalability.
%has
%\pp{the previous paragraph needs to be structured a bit better - first you can explain what are the different options (capture-no capture, co-SF vs. inter-SF interference etc.}

%\rbs{}
% Initial research in LoRaWAN modeled the protocol as pure Aloha channels for each spreading factor and channel pair {as in} \cite{8030482} {which assumes} spreading factors are perfectly orthogonal and does not take capture effect into consideration. 
% {However, the assumption of perfect orthogonality of spreading factors has been shown to be an unrealistic assumption in \cite{10.1007/978-3-319-67639-5_13}, which shows inter-SF interference can overcome the power of a desired transmission resulting in a reception error.

%The inter-SF interference needs a margin of 16 [dB], on average, to disrupt reception. Such a margin can easily arise in an LPWAN just due to the difference in path loss from the gateway to near and far transmitters.}
%\nr{Considering perfect orthogonality for concurrent transmissions can not evaluate a LoRaWAN in the real scenario as is shown in \cite{10.1007/978-3-319-67639-5_13}, where inter-SF interference also shows an important effect on the performance metric. 
% Most papers have analyzed and modeled the network based on the considering complete orthogonality. rbs: I think we should give more examples of this (put more references for the first paragraph)- so we do not just make a claim
{ %A framework for modelling capture effect using stochastic geometry is introduced in \cite{7803607}.
A mechanism impacting frame reception beyond coverage and the capture effect was identified experimentally in \cite{whencollide}. The SX1276 LoRa transceiver is found to lock onto frames after detecting four symbols of a preamble. The reception will fail if another frame (even  one that captures the channel) begins transmission before the first LoRa frame has been received or certain other timing conditions have been fulfilled.

%\nr{Different to previous analytical modeling works in LoRaWAN, Rbs: I'm not sure if we should leave this in or out
{In this paper, we evaluate the scalability of LoRaWAN analytically by developing a joint model for coverage, capture effect and demodulation {capabilities} for LoRa transmissions. We build and evaluate our model on the basis of a SX1301 gateway; however, the model and the methodology can be {applied for any architecture. }%extended to arbitrary number of demodulation paths. 
The SX1301 is capable of demodulating 8 frames, simultaneously, any transmissions beyond this will be dropped. To our knowledge, this practical limitation has not been included in the previous models from the literature.} 
%\nr{ We evaluate the SX1301 chip based gateway which is based on the 8 modulation paths and has an important effect on the performance evaluation where can not demodulate more than 8 packets, simultaneously. This has an important effect but is not evaluated in the previous papers.} 
Finally, we evaluate the performance of several SF allocation schemes based on the developed model.} %\pp{Here you can use the formulation I have in the abstract, that the model is inspired by the Sx1301 chipset, but it has a wider usage.}

The rest of paper is organized as follows: In Section \ref{sec:scenario}, we explain the scenario, assumptions, conditions and parameters of our analysis.  The analysis is contained within Section \ref{sec:analysis}. %The analysis is threefold; we first analyze a collision between two transmissions, then we take FEC into account and finally we consider collisions involving multiple transmissions. 
 Section \ref{sec:results} presents the numerical results and the associated discussion. Finally, the paper is concluded in Section \ref{sec:conclusion}.  

%\nr{\section{NUMERICAL RESULTS and DISCUSSIONS} \label{sec:results}}

\begin{figure}
\centering
 \includegraphics[width=.95\columnwidth]{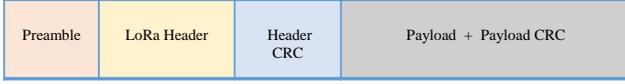}
\caption{LoRa Frame structure as described in \cite{sx1276}. LoRaWAN frames, consisting of header and payload, are transmitted as the payload of a LoRa frame.} %\jjn{Maybe indicate LoRaWAN frame in figure?}} 
    \label{fig:LoRaframe}
\end{figure}

%\nr{{LoRaWAN frames,} consisting of header and payload, are transmitted as the payload of a LoRa frame.}}

%\begin{figure}
%\centering
 %   \includegraphics[width=.95\columnwidth]{LoRaFrame}
  %  \caption{LoRa Frame structure as \cite{sx1276}. {LoRaWAN frames,} consisting of header and payload, are transmitted as the payload of a LoRa frame.}
 %   \label{fig:LoRaframe}
%\end{figure}

\section{Scenario} \label{sec:scenario}

We consider a single LoRa channel with a bandwidth $BW=125$ [kHz]. Each device transmits LoRaWAN frames with payloads of size $B$ bytes at a rate $\lambda_d$. 
A device is assigned a fixed SF $m$ to be used for transmission, where $M = \{7,8,...,12\}$ and $m \in M$ as allowed in LoRaWAN \cite{loraspec2017}. The symbol period for a given $m$ is $T_{s_m} = \dfrac{2^m}{BW}$. 
%\pp{decide on the notation - will you say {$SF=m$} or just \rbs{SF $m$}} 
%fixed spreading factor (SF) 
A LoRA frame uses the LoRaWAN frame as a payload.
Each LoRa frame contains a preamble of $10$ to $65539$ symbols (default is 12), a LoRa header, payload and, optionally, a CRC for the payload as {depicted in}~Fig.~\ref{fig:LoRaframe}.
The number of symbols needed for transmission of the LoRa header and the LoRaWAN payload {using} SF $m$ can be found as \eqref{eq:payloadsymbols} from \cite{sx1276}.
\begin{align} \label{eq:payloadsymbols}
{n_{pkt_m} = 8+\max\Big{(}\Big{\lceil} \dfrac{(8B-4m + 44)}{(4m -2 \cdot \mathbb{I}_\text{DE}) } \Big{\rceil} \cdot (CR+4),0 \Big{)}}{,}
%\color{red}{n_{payload} = 8+max(ceil[\dfrac{(8B-4m + 28+16CRC)}{4(m -2DE) }]\cdot (CR+4),0)\color{red}{,}}
\end{align}
where $CR$ is the coding rate. $CR = i$ for $i\in \{1,2,3,4\}$ corresponds to a coding rate equal to $\frac{4}{4+i}$. The indicator 
$\mathbb{I}_\text{DE}=1$ if low data rate optimization is used and $\mathbb{I}_\text{DE}=0$ otherwise. Low data rate optimization increases robustness towards clock drift and is mandatory for SF \{{$11,12$}\}.
%and is mandatory for SF $m=11,12$.

The total number of symbols transmitted for a complete frame with SF $m$ is $n_{f_m} = n_{pr{e}_m} + n_{pkt_m}$ 
{where $n_{pr{e}_m}$ denotes the number of symbols in the preamble for $m$.} Then, the total transmission time is $T_{f_m} = T_{s_m} \cdot n_{f_m}$.
%\nr{The SX1301 digital baseband chip contains 10 programmable reception paths. }
%\nr{where $n_{frame_m}$, $n_{preamble}$ and $n_{pkt}$ denote the number of symbols per frame m, preamble and packet, respectively.} Then, the total transmission time can be found as $T_{frame_m} = T_{s_m} \cdot n_{frame_m}$ 

%\nr{\subsection{LoRaWAN Deployment}}
\subsection{Deployment Model}
{We consider} $N$ devices distributed %{in a \pp{known manner -PP: what does this mean, are the positions known?} 
{uniformly} in a {circular region with} radius $R$ with a gateway at the center. The distribution of distance $r$ between the gateway and the devices is defined as $g_R(r)$
%\rbs{If the devices are distributed uniformly, the probability density function of device placement to the distance to the gateway $r$ can be expressed as}
{
\begin{equation}
 g_R(r) =\begin{cases}\frac{2r}{R^2} & 0\leq r \leq R\\0 & r \geq R\end{cases} 
\end{equation}
}
The aggregated arrival rate of all devices can be computed as $\lambda=N\cdot \lambda_d$ as long as $\lambda_d < \mu_d$  {where} $\mu_d$ is equal to $\dfrac{1}{{T_{f_m}}/DC}$ and $DC$ is the duty-cycle {given} as a fraction  \cite{7954020}.
%A transmission may be received by multiple gateways in LoRaWAN. %\nr{ We consider both the case of a single receiving gateway and multiple receiving gateways.}

%\begin{figure}
%\centering
 %   \includegraphics[width=.9\columnwidth]{deployment}
 %   \caption{LoRaWAN Deployment}
 %   \label{fig:deployment}
%\end{figure}

%\jjn{In our deployment, we consider gateways to be based on the SX1301, which is {a} popular Digital Baseband Chip for outdoor LoRaWAN macro gateways.  SX1301 is capable of detecting preambles for all spreading factors for up to 8 channels at all times.  However, it is only able to demodulate 8 packets simultaneously as it has 8 parallel demodulation paths. (some repetition)}
Our model is based on SX1301 gateways, which is in widespread use for outdoor LoRaWAN deployments.
SX1301 is capable of detecting preambles for all SFs for up to $8$ channels at a time. This discrepancy between detection and demodulation is due to the fact that the SX1301 architecture separates the preamble detection from the data acquisition \cite{sx1301,sx1301FAQ2018}.
%It is, however, only able to demodulate up to 8 packets simultaneously as it has up to 8 parallel demodulation paths. This discrepancy between detection and demodulation is due to the SX1301 architecture separating preamble detection and data acquisition \cite{sx1301,sx1301FAQ2018}.

\subsection{Channel model}
%\jjn{Missing citations to parameter values in the following.}

%We consider a single end device and evaluate the its uplink PSR regards to co-SF and int-SF interferences.
All devices transmit with power equal to $P_0=14$ [dBm]. 
Let $A(f_c)$ {be} the deterministic part of the path loss model with carrier frequency $f_c$ and noise power $o^2_n = - 174 + 10 \log(BW)$ {[dBm]}, where the noise figure is aassumed to be $0$ [dB]. With the assumed bandwidth, we get $A(f_c)=f^2_c \cdot 10^{-2.8}$. {We define $c$ as in \cite{8480649}
\begin{align} \label{eq:c}
c = \dfrac{{P_0} A(f_c)}{o^2_n}
\end{align}
}
%\pp{$c = \dfrac{{P_0} A(f_c)}{o^2_n}$ - PUT THIS IN NEW LINE AS IT IS REFERRED TO LATER ON} } as {in} \cite{8480649}. 
{We denote the {required} receiver sensitivity for SF $m$ by $\theta_{RX_m}$ and let }
${\gamma}_m^{{{(i)}}}$, ${\gamma}_{co,m}^{{{(i)}}}$ and ${\gamma}_{int,m}^{{{(i)}}}$  {denote the received signal to noise ratio (SNR), the signal to noise plus interference ratios (SINR) of only co-SF interference and only inter-SF interferences for frame $i$, respectively. {${\gamma}_{co,int,m}^{{{(i)}}}$ denotes the received SINR in the presence of co-SF and inter-SF interferences. {${\gamma}_m^{{{(i)}}}$,  ${\gamma}_{co,m}^{{{(i)}}}$, ${\gamma}_{int,m}^{{{(i)}}}$ and ${\gamma}_{co,int,m}^{{{(i)}}}$ are defined as}
%${\gamma}_{co,int,m}^{i}$ denotes the SINR when considering either co-SF or inter-SF transmissions as noise and the other type as interference}. These ratios are defined as the following:
%\nr{and ${\gamma}_{co,int,m}^{i}$ are the received signal to noise ratio (SNR) and signal to noise plus interference ratios (SINR) of the states that there is not any other interference, the co-SF interference, inter-SF interference and the state which there are co and int SFs, respectively.} These ratios are defined as the following:
%${\gamma}_m^{i} , {\gamma}_{co,m}^{i}$ and ${\gamma}_{int,m}^{i}$ \nr{and ${\gamma}_{co,int,m}^{i}$  }are the received signal to noise ratio (SNR) and signal to noise plus interference ratios (SINR) of co-SF interference and int-SF interferences for packet i, {respectively}. These ratios are defined as the following:
%\rbs{comment: c simplifies all the equations why not define it and use it in all the $\gamma$ Eq's?}
%\begin{align}
%&{\gamma}_{m}^{i}=\frac{{p_0}A(f_c) {|h_i|^2} r_i^{-\alpha}}{o^2_n} \nonumber  \\  
%&{\gamma}_{{co,m}}^{i}=\frac{{p_0}A(f_c) {|h_i|^2} r_i^{-\alpha}}{\sum_{k\in k_{co,m}} {p_0}A(f_c){|h_k|^2} r_k^{-\alpha}+o^2_n}   \nonumber \\
%&{\gamma}_{{int,m}}^{i}=\frac{{p_0}A(f_c) {|h_i|^2} r_i^{-\alpha}}{\sum_{k \in k_{int,m} } {p_0}A(f_c){|h_k|^2} r_k^{-\alpha}+o^2_n}
%\end{align}

\begin{eqnarray}
&{\gamma}_{m}^{{(i)}}={c {|h_i|^2} r_i^{-\alpha}} , \\  
&{\gamma}_{{co,m}}^{{(i)}}=\frac{ {|h_i|^2} r_i^{-\alpha}}{\sum_{k\in k_{co,m}} {|h_k|^2} r_k^{-\alpha}+\frac{1}{c}}  ,  \\
&{\gamma}_{{int,m}}^{{(i)}}=\frac{{|h_i|^2} r_i^{-\alpha}}{\sum_{k \in k_{int,m} } {|h_k|^2} r_k^{-\alpha}+\frac{1}{c}} ,\\
&{{\gamma}_{{co,int,m}}^{{(i)}}=\frac{{|h_i|^2} r_i^{-\alpha}}{\sum_{k \in k_{{co},m} } {|h_k|^2} r_k^{-\alpha}+\sum_{k \in k_{int,m} } {|h_k|^2} r_k^{-\alpha}+\frac{1}{c}},}
\end{eqnarray}

%\begin{align}

%\end{align}
where $\alpha$ is the path loss exponent and $h$ is the channel coefficient, which is assumed to be Rayleigh distributed. $k_{co,m}$ and $k_{int,m}$ denote the number of interferer users with the same SF and different SFs, respectively.

%We define $c = \dfrac{\nr{p} A(f_c)}{o^2_n}$ such that the received SNR can be expressed as $\gamma_{\nr{m}} = \dfrac{c\cdot h}{r^\alpha}$

% \rbs{
% \begin{enumerate}
% \item Transmission power
% \item Path loss
% \item Rayleigh Fading
% \end{enumerate}
% }
%\jjn{(missing citation for value)} 
\subsection{Transmission success} \label{sec:tr su}
A transmission must capture the channel in order to be successfully received. The channel is captured if the signal passess the thresholds 
$\Gamma_{m}$, {$\Gamma_{co}$} and $\Gamma_{int,m}$ for capturing the channel with respect to noise, co-SF interference and inter-SF interference, respectively. The probabilities for these events happening are:
\begin{eqnarray}
&\Pr \left(\gamma_{m}^{{(i)}}  > \Gamma_{m}\right),& \\
&{\Pr\left({\gamma}_{{co,int,m}}^{{(i)}} > \max(\Gamma_{co},\Gamma_{int,m})\right)},&\\
&\Pr\left(\gamma_{co,m}^{{(i)}} > {\Gamma_{co}}\right),&\\
&\Pr\left(\gamma_{int,m}^{{(i)}} > \Gamma_{int,m}\right).&
\end{eqnarray}

%\begin{enumerate}
%\item ${\Pr \left(\gamma_{m}^{{(i)}}> \Gamma_{m}\right)}$ 
%\item ${\Pr\left({\gamma}_{{co,int,m}}^{{(i)}}>\max(\Gamma_{co},\Gamma_{int,m})\right)}$
%\item $\Pr\left(\gamma_{co,m}^{{(i)}} > {\Gamma_{co}}\right)$
%\item $\Pr\left(\gamma_{int,m}^{{(i)}} > \Gamma_{int,m}\right)$ 
%\end{enumerate}
In the case where there is no inter-SF or co-SF transmission condition 2) reduces to 3) or 4), respectively. Table \ref{tab:gammathres} lists {$\Gamma_{m}$} and $\Gamma_{int,m}$ for $m\in M$. {Notice that $\Gamma_{co}$ is 6 {[dB]}, which is always larger than $\Gamma_{int,m}$ so condition 2) simplifies to ${\Pr({\gamma}_{{co,int,m}}^{{(i)}}>\Gamma_{co})}$}. %\jjn{How did you get the numbers in Table I?}

%%where $\rbs{\gamma}_{\nr{m}}$, $\rbs{\gamma}_{co,\nr{m}}$ and $\rbs{\gamma}_{int,\nr{m}}$ are the received signal to noise ratio (SNR), signal to noise $+$ interference ratios \nr{(SINR)} of co-SF and int-SF interferences, \rbs{respectively}}.

\begin{table}[bp]
{\caption{Channel capture threshold parameters \cite{8480649}}
\label{tab:gammathres}}
\centering
\begin{tabular}{@{}lccc@{}}
\centering
SF & $\theta_{RX_m}$ [dBm] & $\Gamma_{m}$ [dB] & $\Gamma_{int,m}$ [dB] \\ \midrule
7  & -123	& -6	& -7.5          \\
8  & -126	& -9	& -9          \\
9  & -129	& -12	& -13.5          \\
10 & -132	& -15	& -15          \\
11 & -134.5	& -17.5	& -18          \\
12 & -137	& -20	& -22.5        
\end{tabular}
\end{table}

% If all {these conditions} are fulfilled for a given transmission, {it}
A transmission may be received by the gateway if all the conditions for capturing the channel are fulfilled. However, we must consider the demodulation capability of the receiver {which affects the probability of successful reception}. 
%\jjn{SX1301 has 8 demodulation paths such that it can simultaneously demodulate 8 packets. %where it cannot demodulate more than 8 packets simultaneously.
%A demodulation path is assigned to a packet once four consecutive preamble symbols have been detected.
%A demodulation path is assigned to a packet \nr{if there are less than 8 packets in all the demodulation paths}. 
If all demodulation paths are busy, {then any additional detected frame is dropped.}}
A demodulation path will be assigned to a frame after its preamble has been detected, i.e. four consecutive symbols of the preamble are detected. {The header will then be demodulated and if it is correct, so will the rest of the frame.} %The transmission of a frame that captures the channel when all 8 demodulation paths are busy will result in the reception of all transmissions failing.}
%Each path will lock onto the reception of a frame after detecting 4 preambles, it will then listen for a LoRa header describing the frame format and \nr{then} listen for the frame. 
%So transmitting another frame that captures the channel at any point during ongoing reception of a frame will result in both transmissions failing after being busy of all 8 paths. 
Therefore, the timing of the {transmissions} must be carefully accounted for in the model.

\subsection{SF Allocation}

{SFs} are allocated according to a scheme that is based on annuli, i.e. the radial distance of the device from the BS. Each annuli begins at the radial distance $l_{m-1}$  from the center of the cell and goes to $l_{m}${, such that $l_{12} = R$,} as depicted in Fig. \ref{fig:SFalloc}. Let $\delta_m$ {denote} the fraction of the device population assigned to $m$ and let 
%\pp{How are these lengths selected? Put a pointer to the section where this is done. \nr{We have defined 6 line later}}
%\jjn{$g_m(r)$ be the device density distribution for SF $m$ (Unclear. Is it given in devices/m$^2$?)}. 
{$g_m(r)$ be the device {density} distribution for SF $m$.}
%population
{We define ${\Delta_X}$ as {a set of mapping parameters} ${\Delta_X} = \{\{\delta_7, \delta_8, ... , \delta_{12} \},\{l_7, l_8, ... , l_{12}\},\{g_7, g_8, ... , g_{12}\}\}$ for the SF allocation scheme $X$. }% which ${\Delta}$ denotes the SF allocation set.}
%{The set ${\Delta} = \{\{\delta_7, \delta_8, ... , \delta_{12} \},\{l_7, l_8, ... , l_{12}\},\{g_7, g_8, ... , g_{12}\}\}$ is called SF allocation where}
%\pp{You cannot represent the allocation by a set, but by a mapping.}
{We present three different allocation schemes and compute {$\Delta$} {in Sec. \ref{sec:SF all}}.}

{\begin{figure}
\centering    \includegraphics[width=.75\columnwidth]{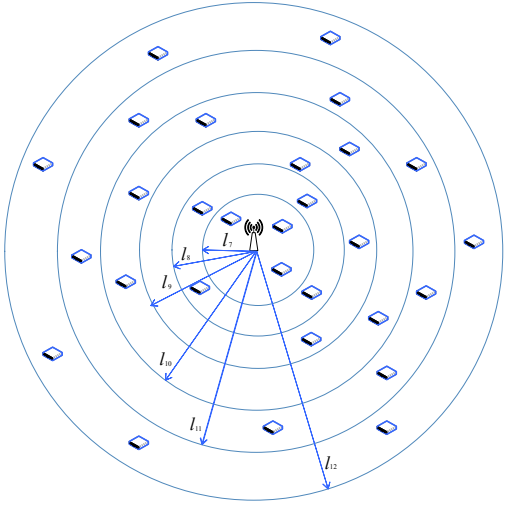}
\caption{{SF} allocation scheme based on annuli in a single gateway LoRaWAN cell.}
 \label{fig:SFalloc}
\end{figure}
}

\section{Uplink performance analysis} \label{sec:analysis}
In this section, we first analyze %the probability of successful reception (PSR) 
%\jjn{FSP (not introduced. Earlier in paper, PSR is mentioned.)\nr{(We can change to FSP and remove all PSR)}} 
the FSP by taking into account the capture effect and the timing of collisions.
%\nr{and the timing of collisions.}
Then, we investigate different SF allocation schemes. % and PSR when multiple gateways are in range.

{ 
A frame is received successfully if it captures the channel and is not dropped due to all demodulation paths being in use. To evaluate the uplink performance, we {derive} the probability of a {LoRa frame from device $i$ being received successfully,} which is denoted by $FSP^{(i)}_m$.
%\jjn{I would suggest to use the notation $FSP^{(i)}_m$, to clearly show that $i$ is an index and does not mean to the power of $i$}.
\begin{align} \label{eq:sucframe}
FSP^{{(i)}}_m ={FCP^{{(i)}}_m} \cdot (1-FDP^{(i)})  ,
\end{align}
%\jjn{Please introduce acronyms. What is the difference between PSR and FSP?} 
%\jjn{Why suddenly talk about successful transmission and not successful reception? The reception is what matters, right?}\\
where  ${FCP^{(i)}_m}$ denotes the probability that the frame captures the channel. ${FDP^{(i)}}$ is the probability a frame being dropped due to all demodulation paths being busy.
}%\\

%
%\nr{
%\begin{align} \label{eq:sucframe}
%P^i_{f_{succes_m}} = (1-P^i_{f_{lock_m}}) \times P^i_{f_{cap_m}}
%\end{align}
%}

%{where ${FLP^i_m}$ and ${FCP^i_m}$ denote the locking and capturing probabilities of channel for the frame $i$ and SF $m$.  }
%\subsubsection{\rbs{Number of interferers}}
%%%%%%%%%%%%%%%%%%%%%%%%%%5

%%%%%%%%%%%%%%%%%%%%%%%%%%5

{To evaluate ${FCP^{{(i)}}_m}$ {and ${FDP^{{(i)}}}$}, we need to {describe} the number of packets which collide with the desired packet. Let $k_{co{,}m}$ and $k_{int{,}m}$ denote the number of co-SF and inter-SF frame transmissions, respectively, which interfere with the transmission of frame $i$ {using} SF $m$. 
{Notice that {we implicitly assume that {all } interfering frames overlap in time by evaluating the total number of {interfering frames} over a an entire frame.} Realistically, several low SF frames could be placed non-overlapping times within the duration of a high SF frame. Therefore, we {derive}  a{ lower bound on the ${FCP^{{(i)}}_m}$.}}
%we are \rbs{evaluating a} {lower bound on the ${FCP^{{(i)}}_m}$.}}
%an upper bound on the interference

We define the traffic load for each source of co-SF and inter-SF interference loads {over a period $\tau$ by $L_{co{,}m}(\tau)$ and $L_{int{,}{m}}(\tau)$, respectively.} When we evaluate the load over a period $\tau$, $L_{co{,}m}(\tau) = (T_{f_m} + \tau)\cdot \lambda \cdot \delta_m$ such that we take into account interference from frame transmissions, which had begun before the start of the observed period $\tau$, but have not ended yet. By equivalent definition, $L_{int{,}{m}}(\tau) =\sum _{p \in M,\\ p\neq m} (T_{f_p} + \tau)\cdot \lambda \cdot \delta_p$ {is the inter-SF load where $p$ denotes all SFs different from $m$.} {Omitting $\tau$ in the notation, the distributions of $k_{co{,}m}$ and $k_{int{,}m}$ are given as:}

\begin{align} \label{eq:Pcolcok}
&P_{k_{co{,}m}} = \dfrac{(L_{co{,}m})^{k_{co{,}m}} \cdot {\exp}(-L_{co{,}m})}{{k_{co{,}m}}!}.
\end{align}
%\rbs{
%\begin{align} \label{eq:Pcolintk}
%&P(k_{int_m} = k) =  \dfrac{1}{\vert M\vert-1} \sum _{p \neq m}\Bigg{(}\dfrac{(L_{int_{m,p}})^k \cdot exp(-L_{int_m})}{k!}\Bigg{)}
%\end{align}
%}
%\nr{And the distribution of $k_{int_m}$ is given by}
\begin{align} \label{eq:Pcolintk}
& P_{k_{int{,}m}} = \dfrac{(L_{int{,}m})^{k_{int{,}m}} \cdot \exp(-L_{int{,}m})}{{k_{int{,}m}}!}.
\end{align}
% \begin{align} \label{eq:Pcolcok}
% &P_{k_{co_m} } = \dfrac{(L_{co,m}^{s})^{k_{co,m} } \cdot exp(-L_{co,m}^{s})}{k_{co,m} !}
% \end{align}
% \nr{\begin{align} \label{eq:Pcolintk}
% &P_{k_{int_m} } =  \sum _{p \neq m}\Bigg{(}\dfrac{(L_{int,m}^{s})^{k_{int,m}} \cdot exp(-L_{int,m}^{s})}{k_{int,m}!}\Bigg{)}
% \end{align}}
%{{Probability of capturing the channel}}
%{With considering an interference-free scenario, the probability of capturing the channel ($CP^i_m $) can be determined as }
%%%%%%%%%%%%%%%%%%%%%%%%%%%%%%%%%%%%%%
\subsection{{Derivation} of ${FCP_m^{{(i)}}}$}
To derive the probability of capturing the channel, 
{we evaluate {the} 4 conditions for {capture effect from} Sec.\ref{sec:tr su}}

%\nr{we evaluate all 4 conditions that are explained in Sec.\ref{sec:tr su}}
%we consider two {cases: No collisions between frames and collision between frames. }

\subsubsection{$k_{co{,}m}+k_{int{,}m}=0 $}
The probability of a single transmission being transmitted in the channel, is the probability of the devices in our deployment not generating a packet within two packet times.
\begin{align}\label{eq:Pcol0}
P_{nocol_m} = \exp\left(-\sum_{p\in M}(T_{f_m} + T_{f_p}) \cdot \delta_p \cdot \lambda \right).
\end{align}
{Then the probability of successfully transmitting the frame is {equal to}}
\begin{align}
{P^{{(i)}}_{s_{nocol}}} =  {P_{nocol_m}} \cdot CP^{{(i)}}_m ,
\end{align}
{where the probability of capturing the channel {is the coverage probability,  $CP^{(i)}_m $, which} can be determined as } 
\begin{align} \label{eq:caprxm}
&CP^{{(i)}}_m = \Pr({\gamma}_{{m}}^{{(i)}} > \Gamma_{{m}}) =\Pr({c|h_i|^2 r_i^{-\alpha}}\geq \Gamma_{{m}}) \nonumber \\ & {=} \int^{l_m}_{l_{m-1}}\exp(- \dfrac{{\Gamma}_{{m}} r_i^\alpha}{c}){g_m(r_i)} dr_i.
\end{align}
{Eq. \eqref{eq:caprxm} can be derived easily based on $|h_i|^2$ being exponentially distributed under our Rayleigh assumption and the PDF of the population to $r_i$ being $g_m(r_i)$.}

\subsubsection{{{$k_{int,m} = 0 $}}}
In the case that there is solely co-SF interference, we can express the probability of capturing the channel as
\begin{equation}
\begin{alignedat}{2} 
&CP_{co,m}^{{(i)}}(k_{co,m}) =\Pr(\frac{ {|h_i|^2} r_i^{-\alpha}}{\sum_{k=1}^{k_{co,m}} {|h_k|^2} r_k^{-\alpha}+\frac{1}{c}} \geq \Gamma_{co}|{k_{co,m}}) \\ & \int^{l_m}_{l_{m-1}}\exp(- \dfrac{{\Gamma}_{co} r^\alpha_i}{c}) {g_m(r_i)} [I(r_i)]^{k_{co,m}}   dr_i ,
\label{eq:capcorx}
\end{alignedat}
\end{equation}
where $I(r_i)$ is equal to
\begin{equation}
\begin{alignedat}{2} 
I(r_i)=\int_{l_{m-1}}^{l_{m}} {\frac{1}{1+{{\Gamma}_{co}}(\frac{r_i}{r})^\alpha}} {g_m(r)}dr.
\end{alignedat}
\end{equation}
\subsubsection{{{$k_{co{,}m} = 0 $}}}
{In this case, there is no co-SF interference, 
$CP^{{(i)}}_{int,m}(k_{int{,}m})$ can be expressed as}
%\nr{In this case where the number of interfere users of co-SF is equal to 0, 
%$CP^i_{int,m}(K)$ can be obtained as}
  
\begin{equation}
\begin{alignedat}{2} 
&CP^{{(i)}}_{int,m}(k_{int{,}m}) =\Pr(\frac{{|h_i|^2} r_i^{-\alpha}}{\sum_{k =1} ^{k_{int,m}}{|h_k|^2} r_k^{-\alpha}+\frac{1}{c}}\geq{\Gamma}_{int,{m}}|{k_{int,m}})\\ &\int^{l_m}_{l_{m-1}}\exp(- \dfrac{{\Gamma}_{int,{m}} r^\alpha_i}{c}) {g_m(r_i)} [\tilde{I}(r_i)]^{k_{int,m}} dr_i, 
\label{eq:capintrx}
\end{alignedat}
\end{equation}

{$\tilde{I}(r_i)$} is defined as
\begin{equation}
\begin{alignedat}{2}
\tilde{I}(r_i)=\int_{{R}\setminus  R_{m}} {\frac{1}{1+{{\Gamma}_{int,{m}}}(\frac{r_i}{r})^\alpha}} {g_p(r)}dr, 
\end{alignedat}
\end{equation}
\subsubsection{{$k_{co{,}m} \cdot k_{int{,}m} \neq 0 $} }
{In case both co-SF and inter-SF interference are present. $CP^{{(i)}}_{int,m}((k_{int{,}m}))$ can be expressed as}
%\nr{In this case where the number of interfere users with the same and different SFs is higher than 1,  $CP^i_{int,m}(K)$ will be}

{  
\begin{equation}
\begin{alignedat}{2} 
&CP^{{(i)}}_{co,int,m}(k_{co,m},k_{int,m}) 
= \\ &pr\Bigg{(}\frac{{|h_i|^2} r_i^{-\alpha}}{\sum_{k =1} ^{k_{co{,}m}}{|h_k|^2} r_k^{-\alpha}+\sum_{j =1} ^{k_{int{,}m}}{|h_j|^2} r_j^{-\alpha}+\frac{1}{c}}
\\&\geq \max({\Gamma}_{co},{\Gamma}_{int,{m}})|{{k}_{int,m}, {k}_{co,m}}\Bigg{)}
\\ &= \int^{l_m}_{l_{m-1}} \exp(- \dfrac{{\Gamma}_{co} r^\alpha_i}{c}) {g_m(r_i)} [I(r_i)]^{{k}_{co,m}}  [{I'}(r_i)]^{{k}_{int,m}} dr_i , 
\label{eq:capintrx}
\end{alignedat}
\end{equation}
}
where $\max({\Gamma}_{co},{\Gamma}_{int,{m}})$ is always equal to ${\Gamma}_{co}$.  ${I'(r_i)}$ is defined as
\begin{equation}
\begin{alignedat}{2}
{I'}(r_i)=\int_{R \setminus R_m} {\frac{1}{1+{{\Gamma}_{co}}(\frac{r_i}{r})^\alpha}} {g_p(r)}dr.
\end{alignedat}
\end{equation}

{Then the probability of capturing the channel can be expressed as a weighed sum of the derived capture probabilities, where the weights are the probabilities of the particular capture scenario taking place. }

{
{\begin{equation}
\begin{alignedat}{2} 
&P_{cap_m} = P_{nocol_m}\cdot CP_m^{{(i)}}
%\nr{ \cdot P_{(k_{int_m}= 0 )} \cdot P_{(k_{int_m}= 0 )}}
\\& + \sum_{k_{co,m} = 1}^{{\infty}}P_{(k_{int,m}= 0 )} \cdot P_{k_{co,m}}\cdot  CP_{co,m}^{{(i)}}({k_{co,m}})
\\& +\sum_{k_{int,m} = 1}^{{\infty}} P_{(k_{co,m}= 0 )} \cdot P_{k_{int,m}} \cdot CP^{{(i)}}_{int,m}({k_{int,m}})
\\ &+\sum_{k_{int,m} = 1}^{{\infty}}\sum_{k_{co,m} = 1}^{{\infty}}P_{k_{co,m}}\cdot P_{k_{int,m}} \cdot CP_{co,int,m}^{{(i)}}(k_{co,m},k_{int,m})
\end{alignedat}
\end{equation}}
}

%\nr{where $\mathbbm{1}_{\{\epsilon\}}$ is the indicator function defined such that $\mathbbm{1}_{\{\epsilon\}} = 1$, if the event $\epsilon$ is true, and 0, otherwise.} 
%{This equation shows $CP^i_{int,m}({k_{int,m}}) $ and $CP_{co,m}^i({k_{co,m}})$ are 1 when $k_{int,m}$ and $k_{co,m}$ are 0, respectively. 
%If $k_{int,m}$ and $k_{co,m}$ are 0, both $CP^{{(i)}}_{int,m}({k_{int,m}})$ and $CP_{co,m}^{{(i)}}({k_{co,m}})$ are set to 1, such that the receiver sensitivity determines reception.} 
{Then, $FCP^{{(i)}}_m$ is given by}

\begin{align} \label{eq:FCP}
FCP^{{(i)}}_m = P_{cap_m}( \tau = T_{f_m} ).
\end{align}

{
%{\begin{equation}
%\begin{alignedat}{2} 
%&P_{cap_m} = P_{nocol_m}\cdot CP_m^i
%\nr{ \cdot P_{(k_{int_m}= 0 )} \cdot P_{(k_{int_m}= 0 )}}
%\\& + \sum_{k_{co_m} = 1}^{\infty}P_{k_{co_m}}\cdot  CP_{co,m}^i({k_{co_m}}) \cdot P_{(k_{int_m}= 0 )} \\& +\sum_{k_{int_m} = 1}^{\infty} P_{k_{int_m}} \cdot CP^i_{int,m}({k_{int_m}}) \cdot P_{(k_{co_m}= 0 )} \\ &+\sum_{k_{int_m} = 1}^{\infty}\sum_{k_{co_m} = 1}^{\infty}P_{k_{co_m}}\cdot P_{k_{int_m}} \cdot min(CP^i_{int,m}({k_{int_m}}),
%\\ &CP_{co,m}^i({k_{co_m}})).
%\end{alignedat}
%\end{equation}}
%}
%\end{comment}

{
%{\begin{equation}
%\begin{alignedat}{2} 
%&P_{cap_m} = P_{nocol_m}\cdot CP_m^i
%\nr{ \cdot P_{(k_{int_m}= 0 )} \cdot P_{(k_{int_m}= 0 )}}
%\\& + \sum_{k_{co,m} = 1}^{\nr{K_{co,m}}}P_{k_{co,m}}\cdot  CP_{co,m}^i({k_{co,m}}) \cdot P_{(k_{int,m}= 0 )} \\& +\sum_{k_{int,m} = 1}^{\nr{K_{int,m}}} P_{k_{int,m}} \cdot CP^i_{int,m}({k_{int,m}}) \cdot P_{(k_{co,m}= 0 )} \\ &+\sum_{k_{int,m} = 1}^{\nr{K_{int,m}}}\sum_{k_{co,m} = 1}^{\nr{K_{co,m}}}P_{k_{co,m}}\cdot P_{k_{int,m}} \cdot min(CP^i_{int,m}({k_{int,m}}),
%\\ &CP_{co,m}^i({k_{co,m}})).
%\end{alignedat}
%\end{equation}}
}

\subsection{Calculation of ${FDP}^{(i)}$ {and {T}hroughput}}

%The characteristic of demodulation paths for a SX1301-based LoRaWAN gateway is described in Sec. \ref{sec:scenario}. A frame will be dropped if all 8 demodulation paths are busy upon detecting the preamble of the frame. We assume that to detect a preamble, the gateway must detect {atleast} 4 consecutive preamble symbols {as found for S1276 in \cite{whencollide}.} 

%{The characteristic of demodulation paths for a SX1301-based LoRaWAN gateway is described in Sec. \ref{sec:scenario}. A frame will be dropped if all 8 demodulation paths are busy. \nr{It is worth noting that the demodulation and preamble detection tasks are separated in SX1301.} \nr{ Therefore, the packet is not dropped because of finite demodulation paths, if the idle demodulation paths are less than 8 paths in the period of receiving the packet \cite{sx1301}. Then, we derive the FDP regards to that all 8 demodulation paths are busy. It can be expressed as }

% \nr{For this case, we will have 5 states, the demodulator lock on 4, 5 , 6, 7 or 8 symbols. Locking on 4 symbols have 5 states. For 5 symbols, we have 4 states. For 6 symbols, 3 states, and for 7 symbols, 2 states. 8 symbols only has one state. Then the $FDP_{m}^{i}$ will be equal to}

{ %We stated in Sec. \ref{sec:scenario} that a
A SX1276 based LoRa receiver will 'lock onto' a frame once it has received 4 preambles as supported by \cite{sx1276,whencollide}. Preamble detection and frame demodulation are separated in the SX1301 and while it is capable of demodulating 8 frames simultaneously as supported by \cite{sx1301,sx1301FAQ2018}, it is able  to detect 48 preambles at once, i.e. a preamble for every {SF} and channel combination.}

{
Since the preamble symbols and the 4 concurrent symbols needed for detection only constitute a small fraction of the total frame size, %\nr{then we only consider the noise effect on the received symbols and} 
%Rbs: we consider noise AND fading, but this is implicit in the "coverage probability" so no need to state it twice
we approximate the capture probability as the coverage probability.
 %\nr{the following assumption is considered as} 
% \rbs{we make the approximation:} \pp{explain in words what are you doing and justify it}
}

{
\begin{align}
CP^{{(i)}}_{{pre}_m} \approx CP^{{(i)}}_m.
\end{align}
}

%We shall disregard that a frame can be successfully received if it begins transmitting during the reception of CRC of the LoRa-Header of a 'locked in' transmission. Sources on the length and timing of this period are scant and the period of the header-CRC is presumably tiny relative to the transmission time of the entire frame we simplify by disregarding this trend. 
{
%We may then compute
 We evaluate the FDP using the CDF of the Poisson distributed number of frames received by the demodulator at any given {instant},
%moment, 
$k_{M} = \sum_{m=7}^{12} k_m$ where $k_m$ denotes the number of frames being received using {$m$}
such that \newline $L_{{M}} = \sum_{m=7}^{12} L_{m}$ where $L_{m} = \lambda \cdot \delta_m \cdot T_{f_m} \cdot CP^{{(i)}}_{{pre}_m} {\cdot (1-FDP)}$
% Supposing that we need to receive 4 preamble symbols, the probability of the SX1301 based gateway being busy and a frame transmission from device i being 'locked out' is:
}

{
\begin{equation} \label{eq:FDP}
\begin{alignedat}{2} 
&FDP^{(i)} =1- \exp(-L_{M}) \cdot \sum_{k = 0}^7 \dfrac{(L_{M})^{k}}{k!}
\end{alignedat}
\end{equation}
}

Eq. \eqref{eq:FDP} can be interpreted as the probability of there not being 7 or less {concurrent frame receptions at the beginning of the reception of a new frame.} $FSP^{(i)}_m $ is now computable as Eq. \eqref{eq:sucframe} by using Eq. \eqref{eq:FCP} and Eq. \eqref{eq:FDP}.

The throughput of the cell can then be calculated as

{
\begin{equation}
T_{\Delta_{Scheme}} = \sum_{m\in M} FSP^{{(i)}}_m \cdot \delta_m \cdot \lambda \cdot B
\end{equation}
}
where the throughput is defined as successfully received {bytes} of LoRaWAN payload per second.

\subsection{{{SF}} Allocation} \label{sec:SF all}
{
In this subsection, we compute $\delta_m$, $l_{m}$ and $g_m(r)$  for $m \in M$ for {three SF allocation schemes.} %\nr{Then, we compare the network performance for each scheme.}
The cell center and edge {for} every scheme are defined as $l_{6} = 0$ and $l_{12} = R$. All parameters for $\Delta$ which are presented in this subsection, can be found in Tab. \ref{tab:sfalloc}.
}

\subsubsection{Uniform}
{
In this scheme, the cell is not divided into annuli and instead SFs are assigned to devices uniformly. {We denote the SF allocation set for this scheme by $\Delta_{Uni}$.} Thus, $\delta_m = \dfrac{1}{6}$ and $g_m(r) = g_R(r)$. In this case, $l_m = {R}$ and $l_{m-1} = 0$ for $m \in M$ is used in the model. 
}
%$l_m = 1 (km)$
%In this scheme $\delta_m$ is easily computed as $\delta_m = \dfrac{1}{len(M)}$ and devices are $l_{m}$ are calculated based on having an equal number of devices within each annuli $N_m = \rho \cdot A_m$. Since the device density, $\rho$, is considered constant throughout the cell, it is equivalent to ensure the area of all annuli are the same.

\subsubsection{Distance}
{
 {We denote the SF allocation set for this scheme by $\Delta_{Dist}$.} We compute the borders of each annulus based on the deterministic path loss and the receiver sensitivity} {such that} $l_m =\left( \dfrac{P_0 \cdot A(f_c)}{\theta_{RX_m}}\right)^{\dfrac{1}{\alpha}}$. Then, we calculate ${\delta}_m$ as a function of the area of a given annulus to the total cell area $\delta_m = \dfrac{N_m}{N} = \dfrac{A_m}{A}$. {${N_m}$ is {the number of}  {devices that are assigned} SF $m$ and ${A_m}$ denotes the area of the annuli $m$.  }
{This means that $g_m(r) = \dfrac{g_R(r)}{\delta_m}$}.

\subsubsection{Equivalent load}
We let the $l_m$ be the same as for $\Delta_{Dist}$. In this allocation scheme{, denoted $\Delta_{Eqload}$}, we calculate $T_{f_m}$ for each {{{SF}} and assign {{{SF}}s to keep the load equal for each SF such that
%$P_m = \dfrac{T_{frame_1}}{\sum_{s=M} \dfrac{frame_1}{frame_s}}$
{$\delta_m = \dfrac{1/T_{f_m}}{\sum_{s=M} {1/T_{f_s}}}$}
.
We keep the device density uniform throughout each annulus, but we assume a higher device density in the annuli closer to the gateway such that $\sum_{m\in M} (\delta_m | l_m) = 1$.

%{
%We consider the cell radius the same and compute annulli based on the receiver sensitivity as previously, but we consider a skewed distribution of devices within the cell, such that devices are distributed uniformly in each anullus, but the device density is higher in anulli closer to the gateway.% to obtain $\delta_m$. %\jjn{(Doesn't this conflict with our previously defined uniform deployment model?)}
%This was not an assumption.
% We assume the device density is constant throughout each annulus, and the density within each annulus is $g_m(r) = \dfrac{g_R(r)}{\delta_m}$.
% }
% that
%{\color{purple}
%(? )

% Second, we will consider is that the device density is equal in all annuli. Thus we must define width of each annulus based on $\delta_m$. $\delta_m = \dfrac{(l_{m-1})^2}{R^2}-\dfrac{(l_{m})^2}{R^2}$.
% So for $m \in M$ we have that $l_{m} = \sqrt[]{R^2 \cdot \delta_m- (l_{m-1})^2}$}

% \subsubsection{Throughput maximization}

% \subsubsection{Cooperative gateways: distance}

\begin{table}[tbp] 
\caption{{SF Allocation schemes}}
\label{tab:sfalloc}
\centering
\begin{tabular}{|lllllll}
\hline
%                         & \multicolumn{8}{l}{SF allocation scheme}                                                                                                  \\ \hline
\multicolumn{1}{l|}{SF}   & \multicolumn{2}{l|}{$\Delta_{Uniform}$}     & \multicolumn{2}{l|}{$\Delta_{Distance}$}    & \multicolumn{2}{l|}{$\Delta_{Eq load}$}  \\ \hline %& \multicolumn{2}{l|}{$\Delta_{Eq load_2}$}     
\multicolumn{1}{l|}{}   & $\delta_m$ & \multicolumn{1}{l|}{$l_m$} & $\delta_m$ & \multicolumn{1}{l|}{$l_m$} & $\delta_m$ & \multicolumn{1}{l|}{$l_m$} \\ %& $\delta_m$ & \multicolumn{1}{l|}{$l_m$} 
\multicolumn{1}{l|}{7}  & 1/6  & \multicolumn{1}{l|}{-}     & .2     & \multicolumn{1}{l|}{.45}     & .47     & \multicolumn{1}{l|}{.45}  \\%   & .47   %  & \multicolumn{1}{l|}{.68}     
\multicolumn{1}{l|}{8}  & 1/6     & \multicolumn{1}{l|}{-}     & .08     & \multicolumn{1}{l|}{.54}     & .25     & \multicolumn{1}{l|}{.54}  \\%    & .25 \\ %   & \multicolumn{1}{l|}{.85}     
\multicolumn{1}{l|}{9}  & 1/6     & \multicolumn{1}{l|}{-}     & .11     & \multicolumn{1}{l|}{.64}     &  .14    & \multicolumn{1}{l|}{.64}   \\%   & .14 \\ %   & \multicolumn{1}{l|}{.93}     
\multicolumn{1}{l|}{10} & 1/6     & \multicolumn{1}{l|}{-}     & .17     & \multicolumn{1}{l|}{.76}     & .08     & \multicolumn{1}{l|}{.76}  \\%    & .08  \\ %  & \multicolumn{1}{l|}{.97}     
\multicolumn{1}{l|}{11} & 1/6     & \multicolumn{1}{l|}{-}     & .19     & \multicolumn{1}{l|}{.88}     & .04     & \multicolumn{1}{l|}{.88}  \\%    & .04  \\  %  & \multicolumn{1}{l|}{.99}     
\multicolumn{1}{l|}{12} & 1/6     & \multicolumn{1}{l|}{-}     & .25     & \multicolumn{1}{l|}{1}     & .02     & \multicolumn{1}{l|}{1}  \\ \hline%    & .02 \\ \hline %   & \multicolumn{1}{l|}{1}     
\end{tabular}
\end{table}

{\section{Results and discussion}
\label{sec:results}}
%\jjn{The results in this section should be presented in a more structured way. Please explain the key point of each figure.}
{We have evaluated CP, FCP, FDP and FSP analytically using Eq. \eqref{eq:caprxm}, Eq. \eqref{eq:FCP}, Eq. \eqref{eq:FDP} and Eq. \eqref{eq:sucframe}, which are derived in Sec. \ref{sec:analysis}. Furthermore, a Matlab-based simulation of the unacknowledged LoRaWAN uplink was implemented and results were simulated. In the simulation, the reception conditions 1-4 are evaluated at the symbol level, whereas the analytical approximation is evaluated at the frame-level {due to \eqref{eq:Pcolcok} and \eqref{eq:Pcolintk}}. This results in the {FCP} approximation being lower-bound. % in Fig. \ref{fig:resProbsuni}, Fig. \ref{fig:resProbsdist} and Fig. \ref{fig:resProbseq}.
{SX1301 is capable of parallel reception on 8 LoRa channels, {$1 \leq N_c \leq 8$}. In this article we have assumed that the frequency of all $N_c$ channels is the same for the evaluation of FCP, which is a fair assumption since all LoRaWAN channels in the %\jjn{800MHz band (please use official name)}
{EU 863-870MHz ISM Band} are fairly close in frequency. The channels are still considered orthogonal to each-other {when evaluating the FCP}. The FDP takes into account traffic on all channels as {discerned LoRa frames} compete for the same demodulation paths.}
%The FCPs of {SF} $m \in \{7,10,12\}$ are plotted in Fig. 
The FCPs of {SF} $ \{7,10,12\}$ are plotted in Fig. \ref{fig:resSFs} for $\Delta_{Dist}$. {Calculating the load contribution on the FDP for each SF $m$, $L_m/L_M$, we get $.59$ for {SF $12$}, $.1$ for {SF $10$} and $.015$ for {SF $7$}.} Although calculated for a specific allocation scheme{,} we can make some general assertions with regards to the impact of different SFs on the reception probability. The higher SFs take up demodulations paths for much longer {time}, which makes {any transmission for SF $m\in M$ more likely to be dropped.} The FCP is also lower for the higher SF frames, which {have} a greater chance of experiencing interference. A higher coverage probability is seen for SF {$7$}, which can be explained by SF $7$ being allocated to devices in a punctured disc around the gateway.
%\nr{because the smaller region is allocated for users with lower SFs, then the co-SF interference for this users will be lower which provide higher $FCP$.}
%\rbs : no this is not true. The coverage is simply better because the min distance is 0. In the dist scheme we would expect the coverage to be the same for all SFs as allocation is based on sensitivity, but the inner ring is a special case. 
% \rbs{A higher coverage probability is seen for SF $m = 7$, but this is explained by the annulus for SF $m = 7$ beginning at the cell center.}}

{{In Fig. %\ref{fig:resProbsuni}, 
\ref{fig:resProbsdist} and Fig. \ref{fig:resProbseq}{,} we observe }%\jjn{where?} 
that the {coverage} probability for $\Delta_{Dist}$ and $\Delta_{Eqload}$ are $.84$ and $.88$, respectively.} 
%{$\Delta_{Uni}$, $\Delta_{Dist}$ and $\Delta_{Eqload}$,} respectively. 
{The slight increase in {coverage} probability from $\Delta_{Dist}$ to $\Delta_{Eqload}$ can be explained by the allocation of more devices in the {SF $7$} annulus. It is also evident that the FDP has a non-negligible effect on the FSP especially when more devices are allocated to higher SFs{, which is the case for $\Delta_{Dist}$.}}

Results for the {$CP$}, $FCP^{{(i)}}_m$ and $FSP^{{(i)}}_m$ are not plotted for the uniform allocation scheme because the frame-based evaluation of interference results in a very low bound for $FCP^{{(i)}}_m$. This is due to the mixture of frames of both high and low SFs in the uniform allocation scheme.

%\rbs{, which  to-be-edited: which provide enough frames that take a long time to be transmitted and enough frames that take little time to transmit, such that much interference is created for a large umber of frames.}.
%the frame-based evaluation of interference results in a very low bound for $FCP^{{(i)}}_m$ due to the mixture of frames of both high and low SFs in the uniform allocation scheme. 

%Numerical evaluation of FDP and CP were found to be as accurate as for the two other allocation schemes and CP was specifically found to be $.68$.}
%The increase in coverage rate from the distance-based allocation scheme to the equivalent-load-based scheme can be explained by devices using lower spreading factors having slightly better coverage as depicted in Fig. \ref{fig:resSFs}.
%\jjn{Please explain figure more carefully.}

% {It is clear that higher spreading factors have a large impact on FLP, which is expected as the longer transmission times will take up a demodulation path in the gateway for longer. This behavior can be seen in Fig. \ref{fig:resSFs} and is also evident in comparison of Fig. \ref{fig:resProbsuni} and Fig. \ref{fig:resProbsdist} to Fig. \ref{fig:resProbseq}.
% }

The throughput of the three allocation schemes are depicted Fig. \ref{fig:resThr}. While the $\Delta_{Dist}$ provides better coverage than $\Delta_{Uni}$, it does not provide a much larger throughput on a cell basis since many devices are assigned to higher SFs. The equivalent-load allocation scheme assumes an uneven distribution of devices, which may not be the case often in actual deployments, but we see that the throughput is much larger in this case. This hints to cooperative assignment of SFs between gateways may possibly provide remarkable throughput improvements in LoRaWAN.

% \begin{figure}
% \centering
%      \includegraphics[width=.95\columnwidth]{simRes1}
%     \caption{Probabilities associated with frame capture for $\Delta_{Uni}$, $\Delta_{Dist}$ and $\Delta_{Eq}$ }
%     \label{fig:resProbs}
% \end{figure}

\begin{figure}
\centering
\includegraphics[width=.95\columnwidth]{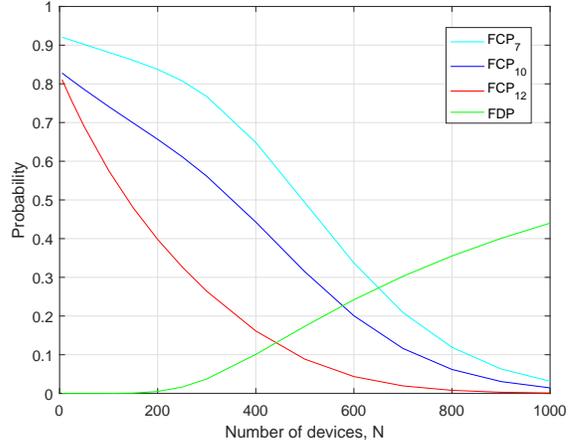}
    \caption{Contribution of {SF}s 7, 10 and 12 on frame capture probability and frame drop probability for $\Delta_{Dist}$. 
    %$\lambda_d = 1/600$, B = 50 and {$N_c = 8$} .
    The traffic model considered was an mean inter-frame generation time of every tenth minute for each device and a payload size of 50 Bytes. This traffic model is assumed to be the same for the{ 8 channels, which are considered.}
    }
    \label{fig:resSFs}
\end{figure}

% \begin{figure}
% \centering
%      \includegraphics[width=.95\columnwidth]{simUniformRes2}
%     \caption{Probabilities associated with frame capture for $\Delta_{Uni}$. 
%     \newline$\lambda_d = 1/600$, B = 50 and {$N_c = 8$} .
%     %The traffic model considered was an mean inter-frame generation time of every tenth minute for each device and a payload size of 50 Bytes. This traffic model is assumed to be the same for the{ 8 channels, which are considered.}
%     }
%     \label{fig:resProbsuni}
% \end{figure}

\begin{figure}
\centering
     \includegraphics[width=.95\columnwidth]{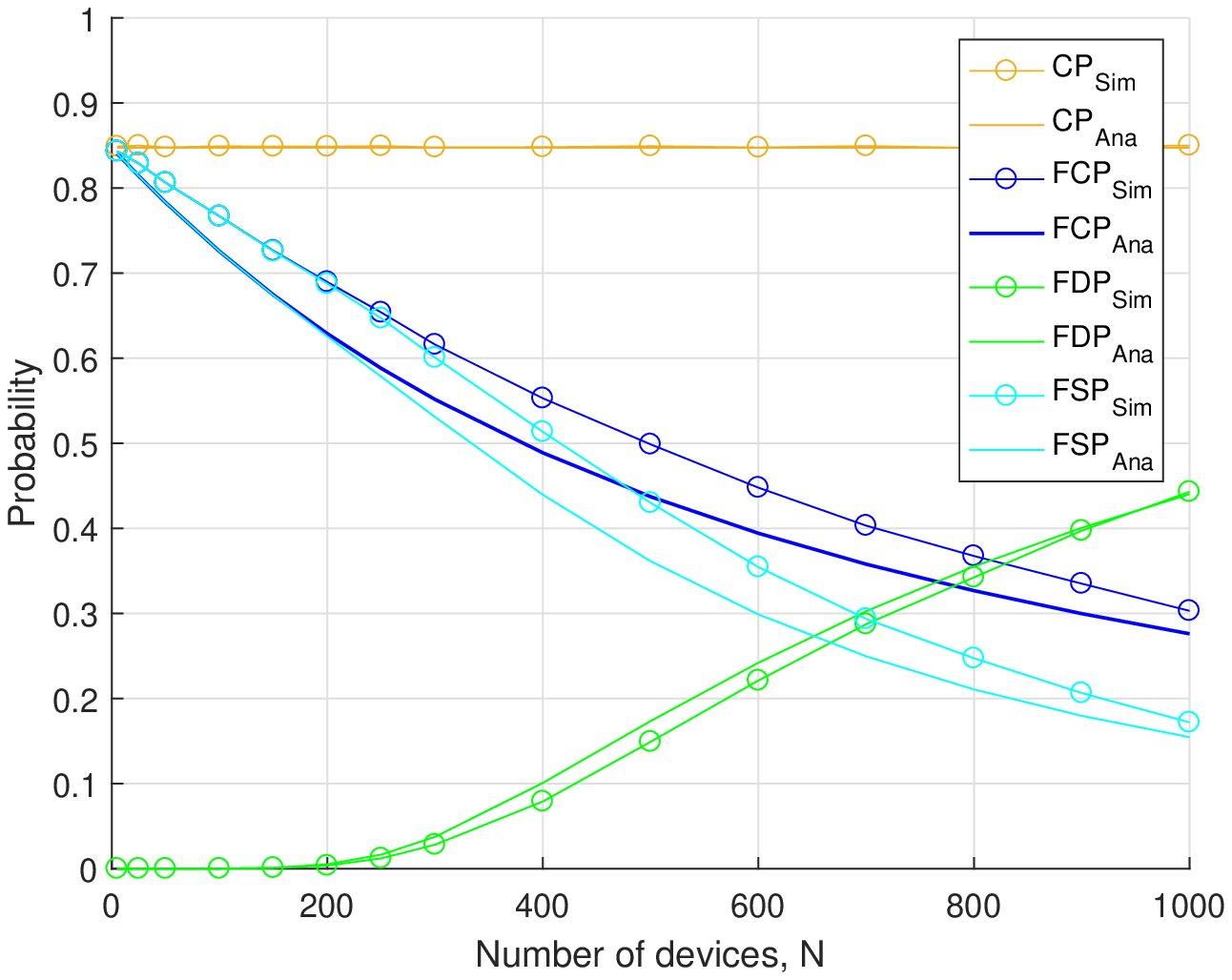}
    \caption{Probabilities associated with frame capture for $\Delta_{Dist}$.\newline$\lambda_d = 1/600$, B = 50 and {$N_c = 8$} .}
    \label{fig:resProbsdist}
\end{figure}

\begin{figure}
\centering
     \includegraphics[width=.95\columnwidth]{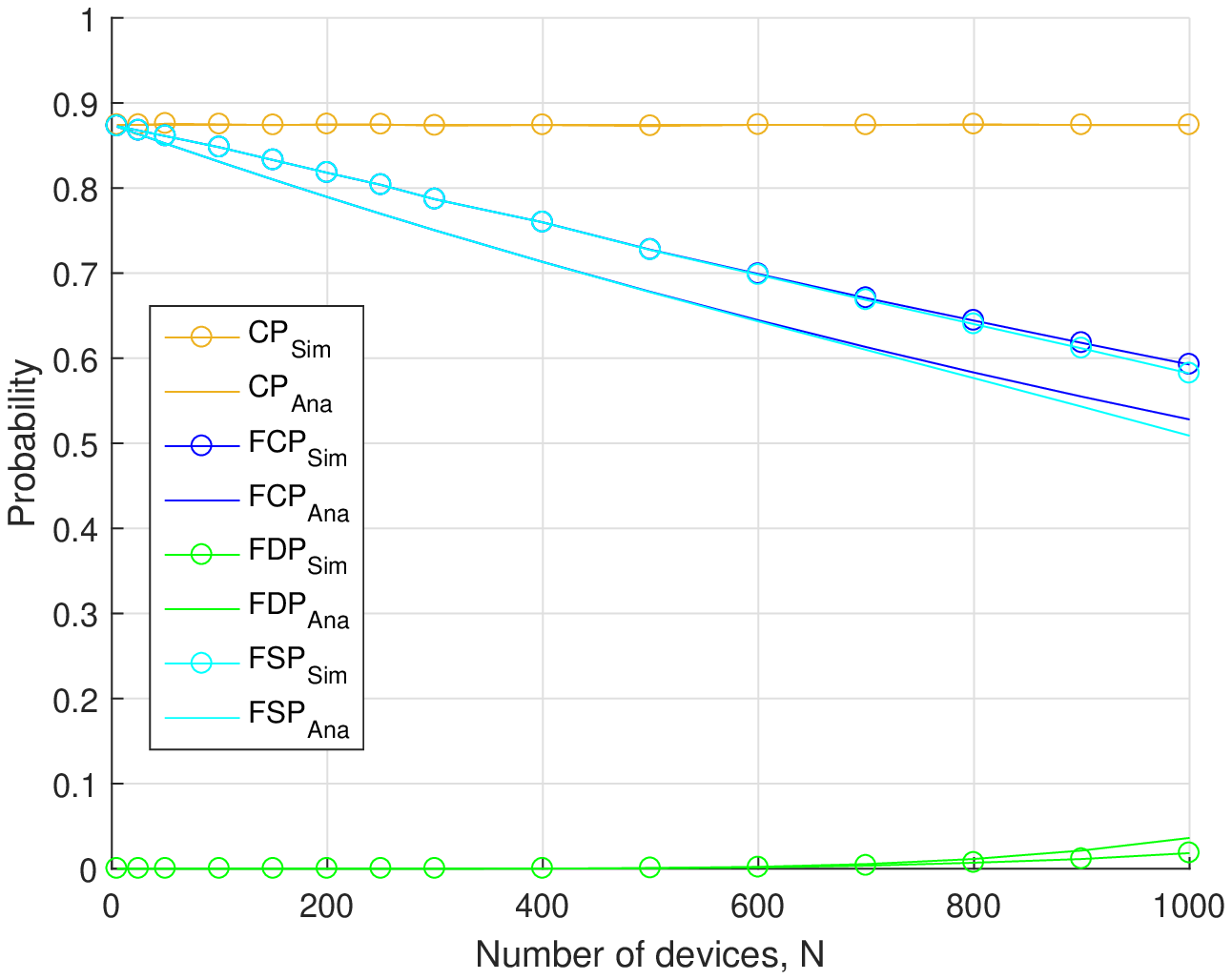}
    \caption{Probabilities associated with frame capture for $\Delta_{Eqload}$.\newline$\lambda_d = 1/600$, B = 50 and {$N_c = 8$} .}
    \label{fig:resProbseq}
\end{figure}

\begin{figure}
\centering
     \includegraphics[width=.95\columnwidth]{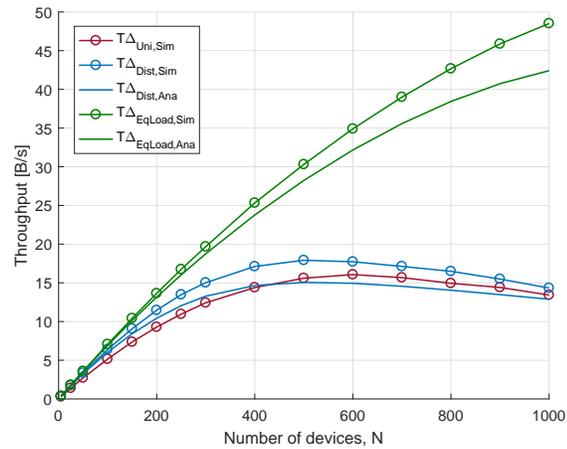}
    \caption{Throughput for the LoRaWAN payload [B/s]. \newline $\lambda_d = 1/600$, B = 50 and {$N_c = 8$} .}
    \label{fig:resThr}
\end{figure}

\vspace{0.5cm}

\section{Conclusion} \label{sec:conclusion}
We have evaluated the uplink performance of a single gateway LoRaWAN deployment in terms of coverage, frame reception success probability and throughput. {Collisions} are evaluated in the time domain based on a traffic model in contrast to other works on capture effect in LoRaWAN, which evaluate capture effect for a fixed number of concurrent transmissions. Unlike previous works, we have considered the demodulation capabilities of the gateway and specifically evaluated SX1301, although the model is applicable to any chipset. 
{We showed that the demodulation capabilities of the receiver have a large impact on the probability of receiving frames successfully.}
%\nr{We have shown that, considering demodulation capabilities have an incredible effect on the frame success probability. } 
%Collisions are evaluated in the time domain based on a traffic model in contrast to other works on capture effect in LoRaWAN, which evaluate capture effect for a fixed number of concurrent transmissions. 
%Unlike previous works, we have considered the demodulation capabilities of the gateway and specifically evaluated SX1301, although the model is applicable to any chipset. Collisions are evaluated in the time domain based on a traffic model in contrast to other works on capture effect in LoRaWAN, which evaluate capture effect for a fixed number of concurrent transmissions.

\bibliographystyle{IEEEtran}
\bibliography{ref}

\vspace{12pt}
%\rbs{NB! 6 pages for conference, up to 1 extra if paid for (above 7 = disqualified)}
%\nr{All "packet"s should be changed to "frame"}

%\textbf{TODOS:}
%\begin{enumerate}
%\item \rbs{FSP == PSR - use FSP to be consistent?} \nr{I also think it is better to use only FSP}
%\item \rbs{We will have to use only "\slash cross" or "\slash cdot" in the end. I don't care which - I just used cdot per habit}

%\end{enumerate}

\end{document}